\documentclass[aps,pra,reprint,amsmath,amssymb,superscriptaddress,nobibnotes,twocolumn]{revtex4-1}
\usepackage{graphicx,SIunits}
\usepackage{bm}     % bold math
\usepackage{hyperref} % add hypertext capabitlities
\usepackage{dsfont}    % allows \mathds{}
\usepackage{color}
\usepackage{booktabs}
\usepackage{blindtext}
\usepackage{amsmath}
\usepackage{multirow}

\begin{document}

%\title{Experimental Demonstration for Reference-Frame and Measurement-Device Independent Quantum Key Distribution with a Few Number of States}

\title{Reference-Frame-Independent, Measurement-Device-Independent \\quantum key distribution using fewer quantum states}

\author{Donghwa Lee}
\affiliation{Center for Quantum Information, Korea Institute of Science and Technology (KIST), Seoul, 02792, Republic of Korea}
\affiliation{Division of Nano \& Information Technology, KIST School, Korea University of Science and Technology, Seoul 02792, Republic of Korea}

\author{Seong-Jin Hong}
\affiliation{Center for Quantum Information, Korea Institute of Science and Technology (KIST), Seoul, 02792, Republic of Korea}
\affiliation{Department of Physics, Yonsei University, Seoul, 03722, Republic of Korea}

\author{Young-Wook Cho}
\affiliation{Center for Quantum Information, Korea Institute of Science and Technology (KIST), Seoul, 02792, Republic of Korea}

\author{Hyang-Tag Lim}
\affiliation{Center for Quantum Information, Korea Institute of Science and Technology (KIST), Seoul, 02792, Republic of Korea}

\author{Sang-Wook Han}
\affiliation{Center for Quantum Information, Korea Institute of Science and Technology (KIST), Seoul, 02792, Republic of Korea}
\affiliation{Division of Nano \& Information Technology, KIST School, Korea University of Science and Technology, Seoul 02792, Republic of Korea}

\author{Hojoong Jung}
\affiliation{Center for Quantum Information, Korea Institute of Science and Technology (KIST), Seoul, 02792, Republic of Korea}

\author{Sung Moon}
\affiliation{Center for Quantum Information, Korea Institute of Science and Technology (KIST), Seoul, 02792, Republic of Korea}
\affiliation{Division of Nano \& Information Technology, KIST School, Korea University of Science and Technology, Seoul 02792, Republic of Korea}

\author{Kwangjo Lee}
\affiliation{Department of Applied Physics, Kyung Hee University, Yongin, 17104, Republic of Korea}

\author{Yong-Su Kim}
\email{yong-su.kim@kist.re.kr}
\affiliation{Center for Quantum Information, Korea Institute of Science and Technology (KIST), Seoul, 02792, Republic of Korea}
\affiliation{Division of Nano \& Information Technology, KIST School, Korea University of Science and Technology, Seoul 02792, Republic of Korea}

\date{\today} 

\begin{abstract}
\noindent Reference-Frame-Independent Quantum Key Distribution (RFI-QKD) provides a practical way to generate secret keys between two remote parties without sharing common reference frames. On the other hand, Measurement-Device-Independent QKD (MDI-QKD) offers high level of security as it immunes against all the quantum hacking attempts to the measurement devices. The combination of these two QKD protocols, i.e., RFI-MDI-QKD, is one of the most fascinating QKD protocols since it holds both advantages of practicality and security. For further practicality of RFI-MDI-QKD, it is beneficial to reduce the implementation complexity. Here, we have shown that RFI-MDI-QKD can be implemented using fewer quantum states than those of its original proposal. We found that, in principle, the number of quantum states for one of the parties can be reduced from six to three without compromising security. Comparing to the conventional RFI-MDI-QKD where both parties should transmit six quantum states, it significantly simplifies the implementation of the QKD protocol. We also verify the feasibility of the scheme with the proof-of-principle experiment.
\end{abstract}

%\keywords{Multiparty quantum communication, Entanglement, Information symmetry}

\maketitle

%\section{Introduction}

{\it Introduction.--} Quantum key distribution (QKD) enables information theoretically secure distribution of random bit strings between two remote parties, Alice and Bob~\cite{Bennett84,Ekert91}. Significant effort has been devoted in order to improve the security and the practicality of QKD. Measurement-Device-Independent QKD (MDI-QKD) is one of the most fascinating QKD protocols in terms of the security improvement. Since MDI-QKD is based on measurement induced entanglement, it is resistant from all the quantum hacking attempts on measurement devices including single photon detectors~\cite{braunstein12,Lo12,Choi16,park18}. Note that MDI-QKD can be implemented with reasonable experimental resources, and thus, the implementations of both long distance communication~\cite{yin16} and high secret key generation rate~\cite{comandar16,liu19a} have been reported. Recently, MDI-QKD has been further developed to Twin-Field QKD and its variants which provide much longer communication distance~\cite{luca18,minder19,liu19b,zhong19a,wang19}.

On the other hand, Reference-Frame-Independent QKD (RFI-QKD) is a representative QKD protocol which improves the practicality of QKD communication~\cite{Laing10,wabnig13,zhang14,liang14,tanu18,yoon19}. In ordinary QKD protocols, it is essential to share common reference frames between Alice and Bob. For example, the polarization axes or the relative phase difference of interferometers should be shared between Alice and Bob for free-space QKD using polarization qubits, and fiber based QKD using time-bin qubits, respectively. Sharing common reference frames, however, can sometimes become difficult and costly. For example, it is not easy to maintain the polarization axes in satellite based free-space QKD~\cite{npj17}. It also becomes challenging to share the relative phase references in the QKD network system where multiple parties participate in the QKD communication~\cite{park19}. RFI-QKD loosen the requirement of sharing reference frames. In RFI-QKD, the secret keys are shared via the basis which is not affected by the reference frame rotation and fluctuation while the security is checked by other two non-commuting bases. It is remarkable that the concept of RFI can be applied to MDI-QKD, i.e., RFI-MDI-QKD, which provides implementation practicality and higher communication security at the same time~\cite{yin14,Wang15_1,zhang17a,wang17,zhang17b,liu18,zhang19b}.

For further practicality, it is beneficial to simplify the QKD implementation by reducing the necessary number of quantum states~\cite{islam18}. For instance, BB84 protocol can be implemented only with three states, instead of conventional four states, without compromising the security~\cite{fung06,branciard07,tamaki14}. Recently, it has been reported that RFI-QKD and MDI-QKD can also be implemented with fewer number of quantum states~\cite{Wang15_2,liu19,tannous19,zhou20}. %\textcolor{red}{However, reducing the number of quantum states in RFI-MDI-QKD has not been studied yet.}

In this Letter, we propose RFI-MDI-QKD protocol using fewer quantum states. In particular, we found that, in principle, the number of quantum states for one of the parties can be reduced to three without compromising security. Comparing to the conventional RFI-MDI-QKD where both parties should transmit six quantum states, it significantly simplifies the implementation of the QKD protocol. We also verify the feasibility of the scheme with the proof-of-principle experiment.

%\section{Theory}

%\subsection{Standard RFI-MDI QKD}

{\it Theory.--} Let us begin by introducing MDI-QKD protocol~\cite{Lo12}. Figure 1 shows a schematic diagram of MDI-QKD. Alice and Bob randomly assign bit values (either + or -) in three bases of $X, Y,$ and $Z$ to the optical pulses and transmit them to a third party, Charlie. The three bases and two bit values provide six possible quantum states, $|Z^\pm\rangle=|0\rangle {\rm~and~} |1\rangle$, $|X^\pm\rangle=\frac{1}{\sqrt{2}}(|0\rangle\pm|1\rangle)$, and $|Y^\pm\rangle=\frac{1}{\sqrt{2}}(|0\rangle\pm i|1\rangle)$. Then, Charlie performs the Bell state measurement (BSM) and announces the result. With the BSM result, Alice and Bob generate the sifted keys only when they have sent the optical pulses in the same basis. Note that in standard MDI-QKD, it is essential to share reference frames among Alice, Bob, and Charlie, i.e., $\mathcal{W}_A=\mathcal{W}_B=\mathcal{W}_C$ where $\mathcal{W}\in\{X,Y,Z\}$. Here, the subscripts {\it A, B}, and {\it C} refer to Alice, Bob, and Charlie, respectively. 

%%%%%%%%%%%%%%%%%%%%%%%%%
\begin{figure}[b!]
\includegraphics[width=3.3in]{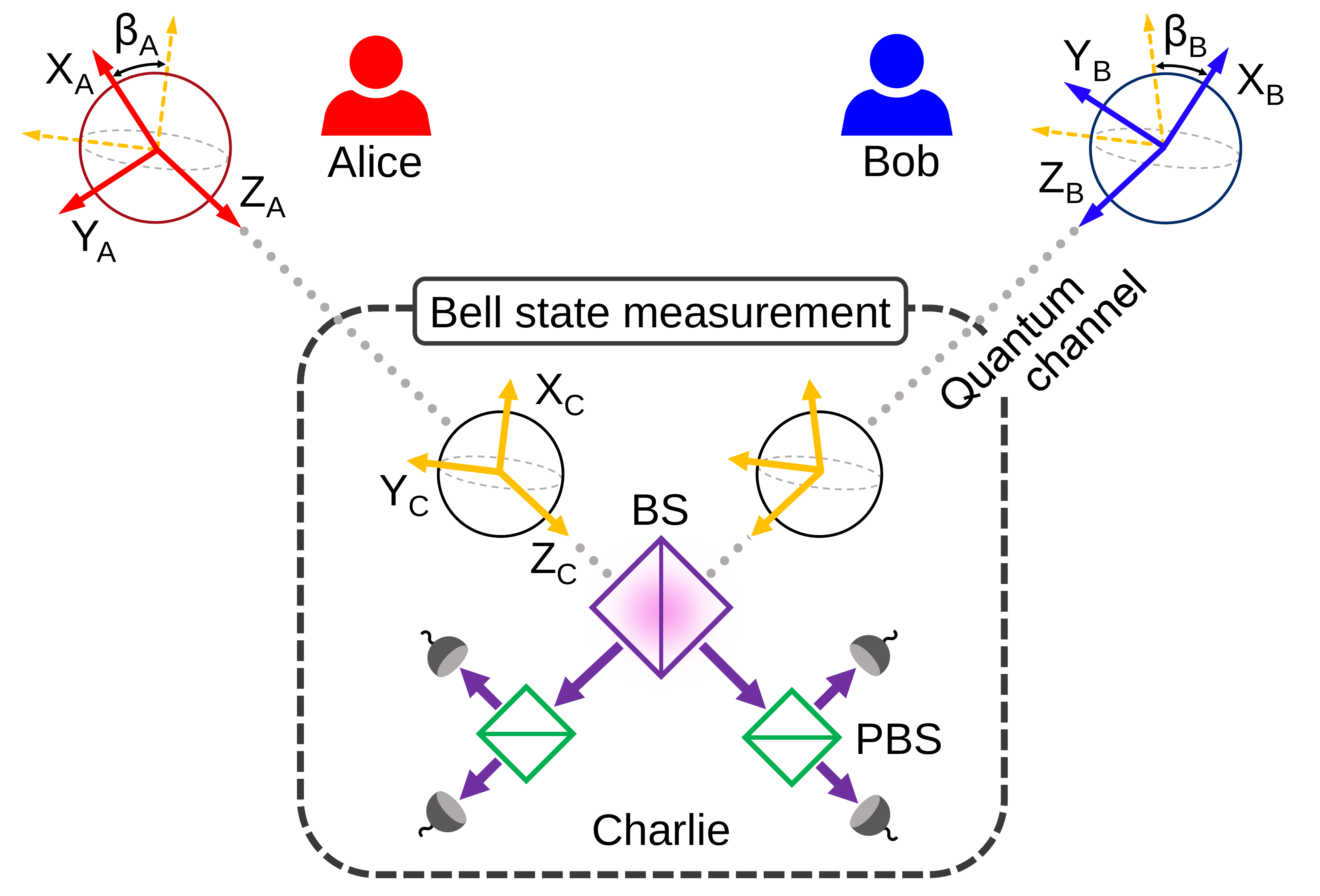} 
\caption{MDI-QKD without shared reference frames. There is the reference frame rotation $\beta_A$ ($\beta_B$) between the frames between Charlie and Alice (Bob).}
\label{concept}
\end{figure}
%%%%%%%%%%%%%%%%%%%%%%%%%

%the eigenstates of $Z$-basis ($|0\rangle$ or $|1\rangle$), $X$-basis ($|\pm\rangle=\frac{1}{\sqrt{2}}(|0\rangle\pm|1\rangle)$) and $Y$-basis ($|\pm i\rangle=\frac{1}{\sqrt{2}}(|0\rangle\pm\ i|1\rangle)$). 
% Note that the linear optical BSM can only detect two Bell states out of four, however, MDI-QKD can be implemented with this incomplete BSM. 

%Each communication subject (Alice, Bob) randomly assigns one of the three possible polarization basis (X, Y, Z) to the optical pulse, then transmits it to a third object (Charlie). The three polarization basis correspond with the six possible states, the eigenstates of $Z$-basis ($|0\rangle$ or $|1\rangle$), $X$-basis ($|\pm\rangle=\frac{1}{\sqrt{2}}(|0\rangle\pm|1\rangle)$) and $Y$-basis ($|\pm\imath\rangle=\frac{1}{\sqrt{2}}(|0\rangle\pm\ i|1\rangle)$). Then, Charlie receives the photon and tries to project them the Bell state $\vert\Psi^\pm\rangle=\frac{1}{\sqrt{2}}(|10\rangle\pm|01\rangle)$, and announces the result, i.e., whether he obtained $|{\rm \Psi}^+\rangle$ or  $|{\rm \Psi}^-\rangle$. With the Bell state measurement result, Alice and Bob generate the sifted key only when they have sent the optical pulses in the same basis. 

Now, let us consider the case when the distant parties do not share common reference frames, but their bases are correlated as Fig.~\ref{concept}~\cite{Laing10}. In this case, each basis has relations of
%%%%%%%%%%%%%%%%%%%%%%%%%%%%%%%
\begin{eqnarray}
Z_C&=&Z_A=Z_B,\nonumber\\
X_C&=&\cos\beta_S X_S+\sin\beta_S Y_S,\label{correlation}\\
Y_C&=&\cos\beta_S Y_S-\sin \beta_S X_S,\nonumber
\end{eqnarray}
%%%%%%%%%%%%%%%%%%%%%%%%%%%%%%%
where $\beta_S$ is the relative reference frame rotation between Charlie and $S\in\{A, B\}$. Note that the bases correlation of \eqref{correlation} can be found in many realistic QKD implementation environment including free space QKD using polarization qubits, and fiber based QKD using time-bin qubits. Since $Z$ basis of all the parties is invariant under the reference frame rotation, Alice and Bob can distribute random bit strings using $Z$ axis. However, the BSM results with the states in $X$ and $Y$ bases depend on the uncontrollable frame rotation $\beta_S$, and thus, the quantum bit error rate (QBER) would increase that limits the performance of the ordinary MDI-QKD.

%%%%%%%%%%%%%%%%%%%%%%%%%%%%%%%
\begin{table}[b!]
\setlength{\tabcolsep}{0.08in}
\centering
\begin{tabular}{|c|c|c|c|c|c|}
\hline
\multirow{2}{*}{Inputs} 	&\multicolumn{2}{c|}{Projector} & \multirow{2}{*}{Inputs}	& \multicolumn{2}{c|}{Projector}\\
\cline{2-3}\cline{5-6}
				& $\langle\Psi_{C}^{+}\vert$ 	& $\langle\Psi_{C}^{-}\vert$ &  & $\langle\Psi_{C}^{+}\vert$ & $\langle\Psi_{C}^{-}\vert$ \\
\hline
$\vert X_A^{+}X_B^{+} \rangle$ & ~~~$\cos\beta$ & $-\cos\beta$ & $\vert X_A^{+}Y_B^{+} \rangle$ & ~~~$\sin\beta$ & $-\sin\beta$\\
$\vert X_A^{+} X_B^{-} \rangle$ & $-\cos\beta$ & ~~~$\cos\beta$ & $\vert X_A^{+} Y_B^{-} \rangle$ & $-\sin\beta$ & ~~~$\sin\beta$\\
$\vert X_A^{-} X_B^{+} \rangle$ & $-\cos\beta$ & ~~~$\cos\beta$ & $\vert X_A^{-} Y_B^{+} \rangle$ & $-\sin\beta$ & ~~~$\sin\beta$\\
$\vert X_A^{-} X_B^{-} \rangle$ & ~~~$\cos\beta$ & $-\cos\beta$ & $\vert X_A^{-} Y_B^{-} \rangle$ & ~~~$\sin\beta$ & $-\sin\beta$\\
$\vert Y_A^{+} Y_B^{+} \rangle$ & ~~~$\cos\beta$ & $-\cos\beta$ & $\vert Y_A^{+} X_B^{+} \rangle$ & $-\sin\beta$ & ~~~$\sin\beta$\\
$\vert Y_A^{+} Y_B^{-} \rangle$ & $-\cos\beta$ & ~~~$\cos\beta$ & $\vert Y_A^{+} X_B^{-} \rangle$ & ~~~$\sin\beta$ & $-\sin\beta$\\
$\vert Y_A^{-} Y_B^{+} \rangle$ & $-\cos\beta$ & ~~~$\cos\beta$ & $\vert Y_A^{-} X_B^{+} \rangle$ & ~~~$\sin\beta$ & $-\sin\beta$\\
$\vert Y_A^{-} Y_B^{-} \rangle$ & ~~~$\cos\beta$ & $-\cos\beta$ & $\vert Y_A^{-} X_B^{-} \rangle$ & $-\sin\beta$ & ~~~$\sin\beta$\\
\hline
\end{tabular}
\label{table}
\caption{The  BSM results with various input states. Here, $\beta=\beta_A-\beta_B$.}
\end{table}
%%%%%%%%%%%%%%%%%%%%%%%%%%%%%%%

In order to solve this problem, RFI-MDI-QKD introduces a new frame rotation invariant parameter $C$, and utilizes it to check the security of the quantum channel. The $C$ parameter is given as~\cite{Laing10,Wang15_1}
%%%%%%%%%%%%%%%%%%%%%%%%%%%%%%%
\begin{eqnarray}
C_{44}={\langle}X_AX_B{\rangle}^2+{\langle}X_AY_B{\rangle}^2+{\langle}Y_AX_B{\rangle}^2+{\langle}Y_AY_B{\rangle}^2.
\label{C_{44}}
\end{eqnarray}
%%%%%%%%%%%%%%%%%%%%%%%%%%%%%%%
Here, the subscript $44$ denotes that both Alice and Bob should transmit four quantum states, $|X^{\pm}\rangle$ and $|Y^{\pm}\rangle$, to estimate the $C$ parameter. Since $|Z^{\pm}\rangle$ states should be transmitted by both Alice and Bob to generate secret keys, the entire RFI-MDI-QKD implementation requires six quantum state transmission for both parties. The expectation value $\langle \mathcal{M}_A\mathcal{N}_B \rangle$, where $\mathcal{M,N}\in\{X,Y\}$, is defined with the BSM results as
%%%%%%%%%%%%%%%%%%%%%%%%%%%%%%%
\begin{eqnarray}
\langle \mathcal{M}_A\mathcal{N}_B \rangle=\dfrac{\pm C^\pm_{++}\mp C^\pm_{+-}\mp C^\pm_{-+}\pm C^\pm_{--}}{\sum_{ij}C^{\pm}_{ij}}.
\label{MN}
\end{eqnarray}
%%%%%%%%%%%%%%%%%%%%%%%%%%%%%%%
Here, $C^\pm_{ij}$ are the BSM results of $\vert\Psi^\pm\rangle=\frac{1}{\sqrt{2}}\left(|01\rangle\pm|10\rangle\right)$ with the inputs of $|\mathcal{M}^i_A\rangle$ and $|\mathcal{N}^j_B\rangle$ where $i,j\in\{+,-\}$. Note that the denominator $\sum_{ij}C^{\pm}_{ij}$ represents the sum of all the BSM results of $\vert\Psi^\pm\rangle$ with all the input states.

With the $C$ parameter and the QBER in $X$ and $Z$ bases, $Q_X$ and $Q_Z$, one can estimate the secret key rate of various QKD protocols. In ordinary MDI-QKD, the secret key rate is given as~\cite{Lo12},
%%%%%%%%%%%%%%%%%%%%%%%%%%%%%%%
\begin{eqnarray}
r_{MDI}=1-H[Q_X]-H[Q_Z],
\end{eqnarray}
%%%%%%%%%%%%%%%%%%%%%%%%%%%%%%%
where $H[x]$ is the Shannon entropy of $x$. In RFI-MDI-QKD, the secrete key rate is given as~\cite{Wang15_1}
%%%%%%%%%%%%%%%%%%%%%%%%%%%%%%%
\begin{eqnarray}
r_{RFI-MDI}=1-H[Q_Z]-I_E[Q_Z,C_{44}],
\end{eqnarray}
where
\begin{eqnarray}
I_E[Q_Z,C_{44}]&=&Q_ZH\left[\dfrac{1+v}{2}\right]+(1-Q_Z)H\left[\dfrac{1+u}{2}\right],\nonumber\\
%\end{eqnarray}
%\begin{eqnarray}
u&=&\min\left[\dfrac{1}{1-Q_Z}\sqrt{\dfrac{C_{44}}{2}},1\right],\\
v&=&\dfrac{1}{Q_Z}\sqrt{\dfrac{C_{44}}{2}-(1-Q_Z)^2u^2}.\nonumber
\end{eqnarray}
%%%%%%%%%%%%%%%%%%%%%%%%%%%%%%%

%\subsection{RFI-MDI QKD using fewer quantum states}

So far, we have shown that the security of RFI-MDI-QKD can be checked with four quantum states sent by both Alice and Bob. Now, we show that the number of quantum states for security check can be reduced by utilizing the correlation of frames \eqref{correlation}. The BSM result probability to obtain $|\Psi^{\pm}\rangle$ is estimated as $|\langle\Psi_C^{\pm}\vert\varphi_A\varphi_B\rangle|^2$ where $|\varphi_A\rangle$ and $|\varphi_B\rangle$ are the input states from Alice and Bob, respectively. We summarize all the possible combinations in $X$ and $Y$ bases at Table~1. We can derive the relations between the expectation values used in \eqref{C_{44}} as
%%%%%%%%%%%%%%%%%%%%%%%%%%%%%%%
\begin{eqnarray}
\langle X_AX_B \rangle &=& \langle Y_AY_B \rangle = \cos\beta,\nonumber\\
\langle X_AY_B \rangle &=& -\langle Y_AX_B \rangle = \sin\beta,
\label{relation2}
\end{eqnarray}
%%%%%%%%%%%%%%%%%%%%%%%%%%%%%%% 
where $\beta=\beta_A-\beta_B$. According to these relations, Alice can omit the $Y_A$ basis in order to obtain the $C$ parameter, i.e.,
%%%%%%%%%%%%%%%%%%%%%%%%%%%%%%%
\begin{eqnarray}
C_{24}=2~\big({\langle}X_AX_B{\rangle}^2+{\langle}X_AY_B{\rangle}^2\big).
\end{eqnarray}
%%%%%%%%%%%%%%%%%%%%%%%%%%%%%%%
Here, the subscript $24$ presents that Alice and Bob, respectively, have to transmit two and four quantum states in order to obtain the $C$ parameter.

It is notable that Alice can further reduce one more state. From Table~1, one can find that 
%%%%%%%%%%%%%%%%%%%%%%%%%%%%%%%
\begin{eqnarray}
\langle \Psi_C^{\pm}\vert \mathcal{M}_A^{+}\mathcal{N}_B^{+} \rangle&=&\langle \Psi_C^{\pm}\vert \mathcal{M}_A^{-}\mathcal{N}_B^{-} \rangle,\nonumber\\
\langle \Psi_C^{\pm}\vert \mathcal{M}_A^{+}\mathcal{N}_B^{-} \rangle&=&\langle \Psi_C^{\pm}\vert \mathcal{M}_A^{-}\mathcal{N}_B^{+} \rangle.
\end{eqnarray}
%%%%%%%%%%%%%%%%%%%%%%%%%%%%%%%
Thus, the $C$ parameter can be estimated with one state from Alice and four states from Bob as 
%%%%%%%%%%%%%%%%%%%%%%%%%%%%%%%
\begin{equation}
C_{14}=2~({\langle}X^{+}_{A}X_B{\rangle}^2+{\langle}X^{+}_{A}Y_B{\rangle}^2).
\label{C_14}
\end{equation}
%%%%%%%%%%%%%%%%%%%%%%%%%%%%%%%
With $C_{14}$, it is sufficient for Alice to send only three states of $|Z_A^\pm\rangle$ and $|X^+_A\rangle$ to perform RFI-MDI-QKD. Note that, $|X^+_A\rangle$ state to estimate $C_{14}$ is not uniquely determined, but any equal superposition of $|0\rangle$ and $|1\rangle$, i.e., $\frac{1}{\sqrt{2}}\left(|0\rangle+e^{i\theta}|1\rangle\right)$ can be utilized. Therefore, it significantly reduces the implementation resources in time-bin encoding RFI-MDI-QKD since Alice does not need a phase modulator to adjust the phase $\theta$.
%\textcolor{red}{We note that either  Alice or Bob}

%$|\langle \Psi_C^{\pm}\vert \mathcal{M}_A^{+}\mathcal{N}_B^{+} \rangle|^2=|\langle \Psi_C^{\pm}\vert \mathcal{M}_A^{-}\mathcal{N}_B^{-} \rangle|^2$. The probability amplitude of Bell projection counts, $C_{ij}^{\pm}$ can be regard as same with projection ratio, $\langle\Psi^{\pm}\vert\varphi_A\varphi_B\rangle$, which is deriving the following relation,
%%%%%%%%%%%%%%%%%%%%%%%%%%%%%%%
%\begin{eqnarray}
%C^{\pm}_{++}=C^{\pm}_{--},~C^{\pm}_{+-}=C^{\pm}_{-+}.
%\end{eqnarray}
%%%%%%%%%%%%%%%%%%%%%%%%%%%%%%%

%%%%%%%%%%%%%%%%%%%%%%%%%
\begin{figure}[b]
\includegraphics[width=3.4 in]{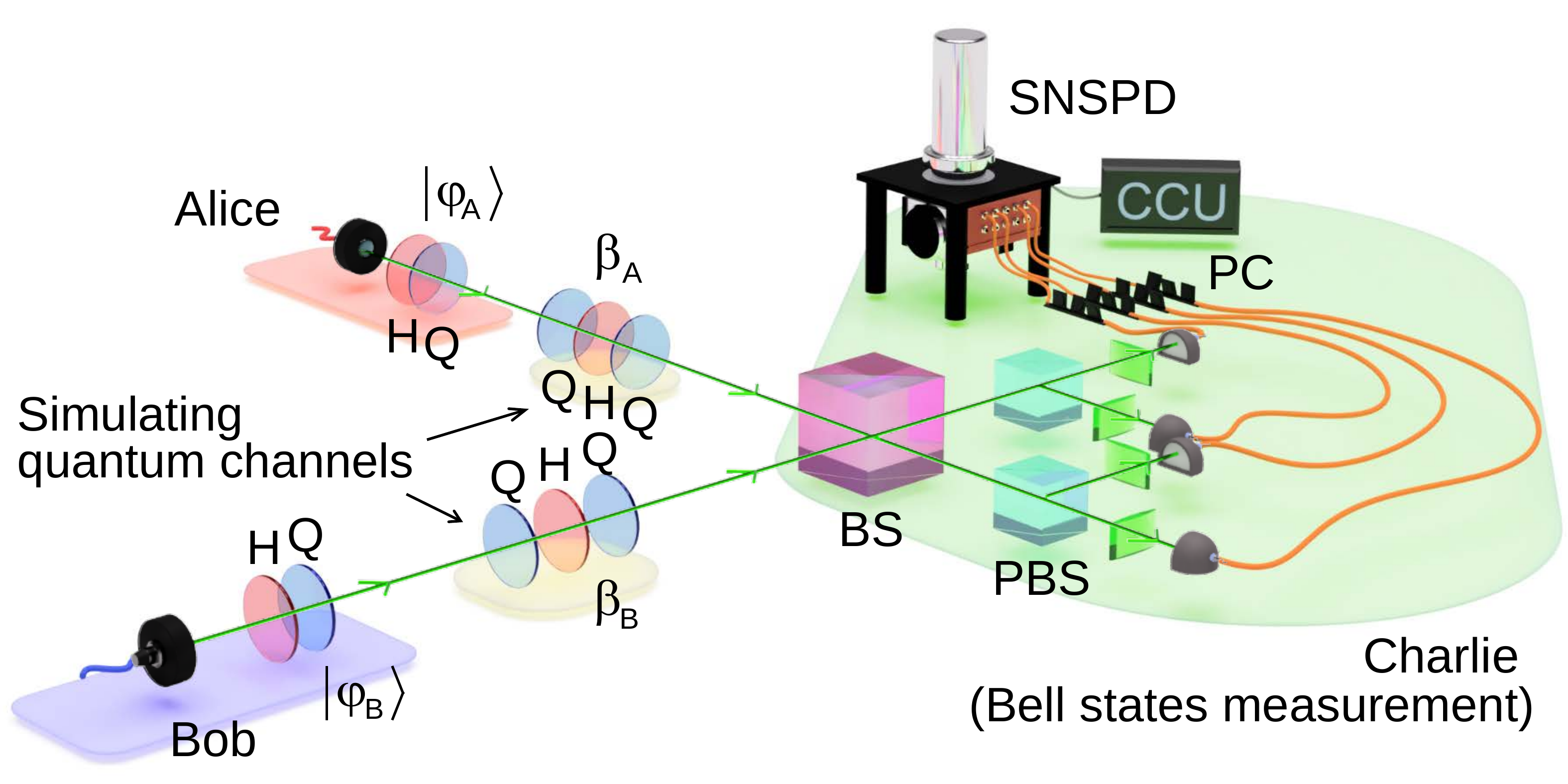} 
\caption{The experimental setup of MDI-QKD and RFI-MDI QKD under reference frame rotation $\beta_A$ and $\beta_B$. H: half waveplate, Q: quarter waveplate, BS: beamsplitter, PBS : polarizing beamsplitter, PC: polarization controller, SNSPD: superconducting nanowire single photon detector, CCU: coincidence counting unit.}
\label{setup}
\end{figure}
%%%%%%%%%%%%%%%%%%%%%%%%%

%\section{Experiment}

%%%%%%%%%%%%%%%%%%%%%%%%%
\begin{figure}[b!]
\includegraphics[width=3.5in]{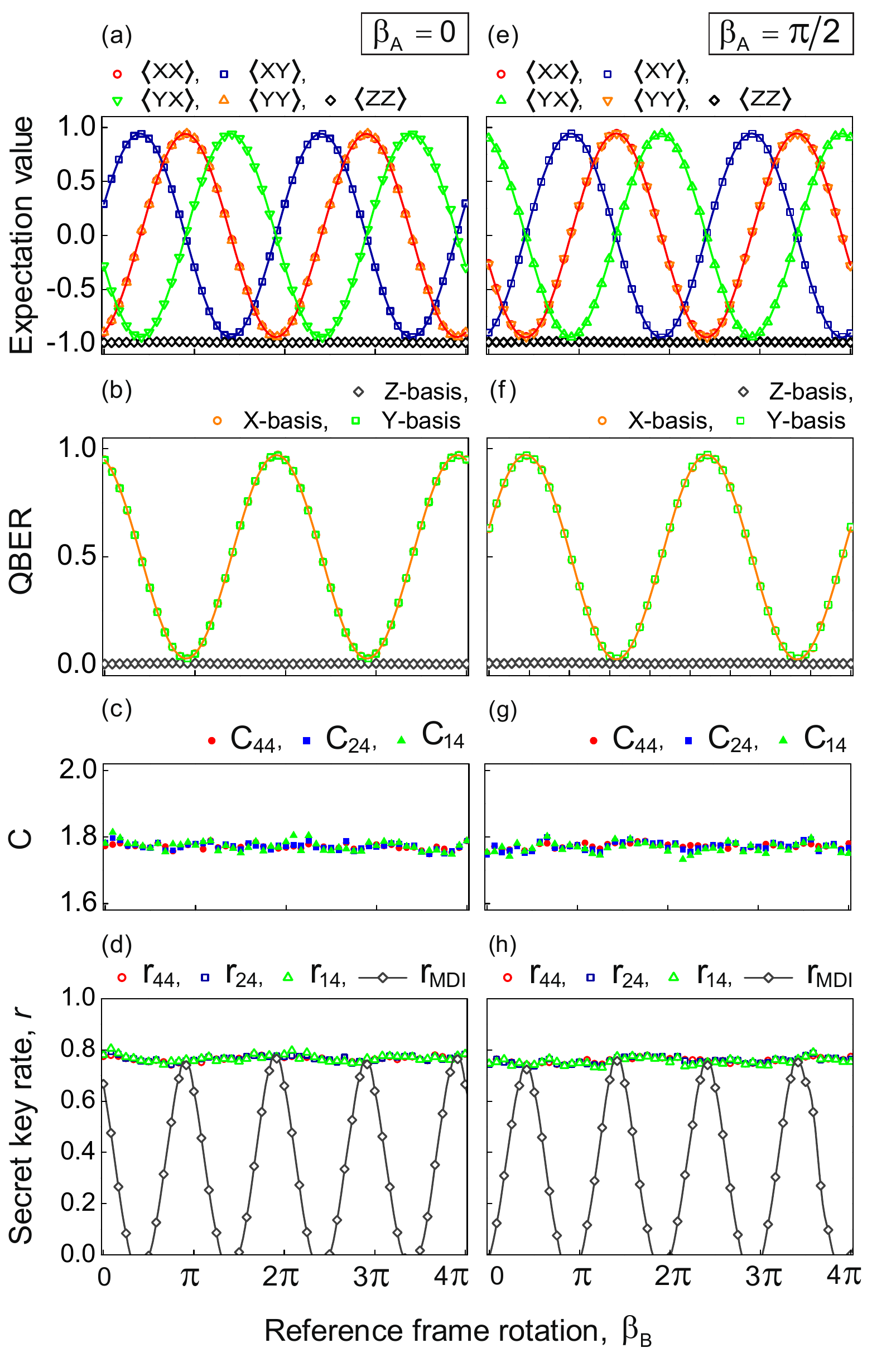} 
\caption{The experimental results of RFI-MDI-QKD. The left and right column are data for $\beta_A=0$ and $\pi/2$, respectively. The experimentally obtained error bars are smaller than the markers. See text for details.}
\label{data}
\end{figure}
%%%%%%%%%%%%%%%%%%%%%%%%%

{\it Experiment.--} Figure~\ref{setup} presents the experimental setup to perform the proof-of-principle experiment on RFI-MDI-QKD using fewer quantum states. The identical single photons at 1556~nm utilized by Alice and Bob are prepared by type-II spontaneous parametric down conversion using 10~mm periodically-poled KTP crystal pumped by femtosecond laser pulses (not shown in Fig.~\ref{setup}). The pump power and repetition rate of the pump are 10~mW and 125~MHz, respectively. Using superconducting nanowire single photon detectors (SNSPDs) with $80\%$ of single-photon detection efficiency, we obtained about 200~kHz coincidences. The indistinguishability of single photons is verified by the Hong-Ou-Mandel (HOM) interference visibility of $V\sim0.99$. Note that the high HOM visibility also implies that the multiple photon generation events are negligible~\cite{takeoka15}.

Then, the random polarization states of $\vert\varphi_A\rangle$ and $\vert\varphi_B\rangle$ are encoded by the sets of half- and quarter-wave plates (H, Q). The single photons are transmitted to Charlie who performs the BSM using a beamsplitter (BS) and polarizing beamsplitters (PBS). The BSM result is obtained by various combinations of two-fold coincidences using a home-made coincidence counting unit (CCU)~\cite{Park15}. In order to precisely simulate the rotating quantum channels, $\beta_A$ and $\beta_B$, we put sets of waveplates in the free-space quantum channel. By rotating a HWP which is placed between two QWPs at $45^\circ$, one can implement the reference frame rotation of \eqref{correlation}~\cite{yoon19}. 

%The light sources used by Alice ans Bob are photon-pair generated by spontaneous parametric down conversion (SPDC) with periodically poled potassium titanyl phosphate (PPKTP) crystal pumped by pulse laser with 8 ns pulse period, and its polarizations are prepared in the arbitrary polarization state $\vert\varphi_A\rangle$, $\vert\varphi_B\rangle$ with a set of half-wave and quarter wave plates (HWP, QWP). Then, in order to exactly implement the polarization axes rotation, we insert a set of wave plates at the free-space channel. By changing the angle, $\theta$ of a HWP which is placed between two QWP at an angle of $\pi/4$, we can demonstrate the polarization axes rotation. With calculating a unitary matrix of set, we know the $\beta_A=4\theta_A$ ($\beta_B=4\theta_B$), one can realize the reference frame rotation of Eq. (1). Then, the two photons are sent to beam splitter (BS) and occur single photon interaction in it. With following a set of polarization beam splitter (PBS) and detector, the Bell basis projection is performed. We use a superconducting nanowire single photon detector (SNSPD) for detecting the photon in Telecom band. Finally we can get Bell projection results from two-fold coincidence following a detector assignment.

In order to investigate the effect of the reference frame rotating quantum channels, we have performed both MDI-QKD and RFI-MDI-QKD with respect to $\beta_B$ with fixed $\beta_A$. The left and right column of Fig.~\ref{data} presents the experimental data for fixed $\beta_A=0$ and $\beta_A=\pi/2$, respectively. Figure~\ref{data} (a) and (e) present various expectation values of $\langle \mathcal{M}_A\mathcal{N}_B \rangle$. While $\langle ZZ \rangle=-~0.987\pm 0.007$ is invariant under the varying $\beta_B$, all other expectation values show the sinusoidal oscillation with the visibility of $V=0.945\pm 0.002$. We note that the sinusoidal oscillations have phase shift from the theoretical values of Table~1 due to the non-ideal phase factor at the Charlie's BS for the BSM. It also verifies the relation of \eqref{relation2}. With the expectation values, one can estimate QBER in various bases, see Fig.~\ref{data}(b) and (f). Note that, while we present the raw data for the QBER for clarity, QBER $>0.5$ can become  QBER $<0.5$ by simply flipping one of the bit values. The QBER in $X$ and $Y$ bases show sinusoidal oscillation with respect to $\beta_B$, while that of $Z$ basis is unchanged. The experimentally obtained QBER in $Z$ basis is $Q_Z=0.7\pm 0.1\%$. The $C$ parameters are shown in Fig.~\ref{data}(c) and (g). It clearly shows that the $C$ parameters are invariant under the varying $\beta_B$. More significantly, all the $C$ parameters of $C_{44}$, $C_{24}$, and $C_{14}$ are very similar to each other. In particular, we found that $C_{44}=1.77\pm 0.03$, $C_{24}=1.77\pm 0.02$, and $C_{14}=1.77\pm 0.03$. Finally, we have presented the secret key rate $r$ in Fig.~\ref{data}(d) and (h). It shows that the secret key generation rate of an ordinary MDI-QKD varies with $\beta_B$, and fails to generate positive secret keys in some regions. On the other hand, RFI-MDI-QKD can always generate the secrete keys with any $\beta_B$. It also shows that the secret key rates of $r_{44}$, $r_{24}$, and $r_{14}$ are very similar meaning that the performance of RFI-MDI-QKD using fewer quantum states is comparable.

It is noteworthy that the security analysis is valid for a small $Q_Z$. A large $Q_Z$ can limit the theoretical upper bound of the $C$ parameter, and thus, may affect to the performance of the QKD with fewer quantum states~\cite{liu19}. We remark, however, our security analysis shows the benefit of using fewer quantum states in real world implementation. In RFI-MDI-QKD, $Q_Z$ is obtained by the simple projection measurements onto separable states. On the other hand, the C parameter is estimated with the interferometric results in $X$- and $Y$-bases. Therefore, $C$ parameter is much more fragile than $Q_Z$ with noise and experimental imperfections, e.g., see our experimental results and Ref.~\cite{wang17}. This suggests that the security of practical RFI-MDI-QKD would be limited by the experimentally obtainable $C$ parameter rather than the theoretical upper bound. Our results show that the experimentally obtainable $C$ parameters are independent of the number of quantum states, and thus, the performance of RFI-MDI-QKD using fewer quantum states would be comparable.

% \section{Conclusion}

{\it Conclusion.--} In conclusion, we have proposed RFI-MDI-QKD using fewer quantum states. In particular, we have verified that transmitting only three quantum states from one of the transmitters is sufficient for implementing RFI-MDI-QKD. Considering the ordinary RFI-MDI-QKD protocol requires that both two parties transmit six quantum states, it significantly simplifies the implementation of the QKD protocol. We remark that, in optical fiber based QKD using time-bin qubits, the three quantum states transmission does not require a phase modulator to adjust the phase of the interferometer, and thus, it significantly reduces the implementation complexity. The effectiveness and feasibility of our scheme have also been verified with proof-of-principle experimental demonstration. Our results indicate that, in real world implementation of QKD in earth-to-satellite links or chip-to-chip integrated photonic waveguides, the RFI-MDI-QKD protocol can be a simpler but more powerful solution for a better QKD communication. We remark the security analysis with more practical systems, e.g., weak coherent pulse inputs and finite key analysis, would be interesting future works.

%\section*{Data availability}
%The data that support the findings of this study are available from the corresponding authors upon reasonable request.

\section*{Acknowledgement}
This work was supported by the NRF programs (2019 M3E4A1079777, 2019R1A2C2006381, 2019M3E4A107866011), and the KIST research program (2E30620).

%\section*{Disclosures}
%The authors declare no conflicts of interest.

\end{document}